      [1999/12/01 v1.4c Il Nuovo Cimento]
\documentclass{cimento}

\newcommand{\ltsima} {$\; \buildrel < \over \sim \;$}
\newcommand{\gtsima} {$\; \buildrel > \over \sim \;$}
\newcommand{\lta} {\lower.5ex\hbox{\ltsima}}
\newcommand{\gta} {\lower.5ex\hbox{\gtsima}}

\newcommand{\be}{\begin{equation}}
\newcommand{\ee}{\end{equation}}
\newcommand{\ba}{\begin{eqnarray}}
\newcommand{\ea}{\end{eqnarray}}
\newcommand{\siml}{\lower4pt \hbox{$\buildrel < \over \sim$}}
\newcommand{\simg}{\lower4pt \hbox{$\buildrel > \over \sim$}}
\usepackage{graphicx}  % got figures? uncomment this
\title{ESTREMO: Extreme phySics in the TRansient and Evolving COsmos}
\author{E.~Costa\from{ins:x} \atque
L.~Piro\from{ins:x} \thanks{on behalf of the ESTREMO
collaboration}} \instlist{\inst{ins:x}Istituto Astrofisica
Spaziale e Fisica Cosmica - Rome, INAF }
\PACSes{\PACSit{95.85.Nv}{X--rays}
\PACSit{98.70.Rz}{gamma--ray sources, gamma--ray bursts}}
\begin{document}

\maketitle

\begin{abstract}
We present a mission designed  for the study of transient
phenomena in the high energy sky, through a wide field X--ray/hard
X--ray monitor, and fast ($<$ 1 min) follow up observations with
Narrow Field Instrumentation. This is based on an X--ray telescope
with an area of 1000 cm$^2$, equipped with high­-resolution
spectroscopy microcalorimeters and X--ray polarimeter. The
performances of the mission on the physics of GRB and their use as
cosmological probes are presented and discussed.
\end{abstract}

\section{Scientific goals}
The ESTREMO mission is devoted to address, through a smart,
focussed approach, two main themes of Astrophysics and Cosmology:
\begin{itemize}
\item The study of extreme objects in our Universe, in particular
those characterized by very large energy release over short time
scale (minutes-hours) such as Gamma-ray Bursts, Massive and
star-size black Holes, Neutron stars, Supernovae explosions,
flaring sources.
 \item The study of the Evolution of the Universe
by using the brightest and most distant explosions, the Gamma-Ray
Bursts, as distant beacons.
\end{itemize}

\section{Mission Profile}

The mission is designed to localize transient sources in outburst
and to observe them while they are still in outburst phase.
Catching a source during large and fast flare activity gives
access to the study of very large energy release under extreme
conditions. In addition, in this phase the flux can be orders of
magnitude greater than in quiescent state, thus allowing very
sensitive observations with relatively  small effective area. The
mission comprises:

\begin{enumerate}
\item a wide field instrument, to localize X-ray transient
phenomena in the sky in the X-ray and hard X-ray range (2-300
keV): as a baseline an array of CdZnTe with coded mask
 \item an autonomously fast pointing (1 min) satellite for
 follow-up observations by narrow-field instruments with an X-ray telescope
 of $1000$ cm$^2$
\end{enumerate}
to perform:
\begin{itemize}
\item high resolution X-ray spectroscopy with TES (Transition Edge
Sensor) microcalorimeters (Resolution $\approx$ 2-4 eV in the
0.1-10 keV range) \item High sensitivity X-ray polarimetry
\end{itemize}
Auxiliary instrumentation   should extend the spectral coverage of
prompt emission to soft X-rays and to the MeV region. To underline
the mission capabilities with regard to GRB, we present two
examples, one based on high-resolution spectroscopy, the other on
polarimetry.

\begin{figure}
%\centerline{\psfig{figure=f1e-5_r4_whim.ps,angle=-90,width=17cm,height=9cm}}
\includegraphics[width=9cm,origin=c,angle=-90]{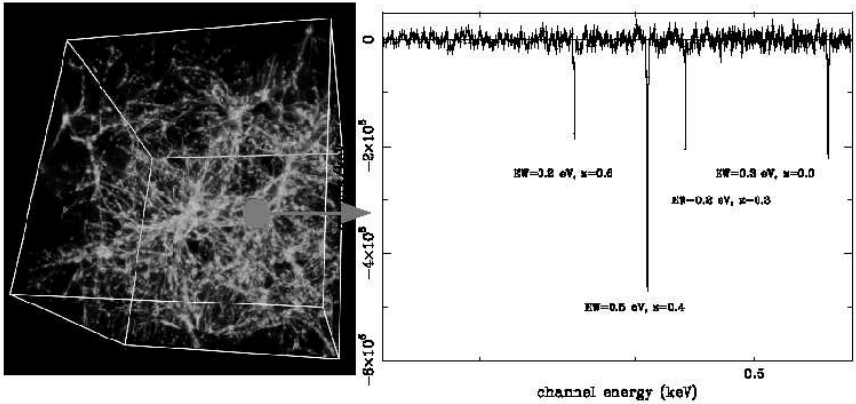}
\vspace{-3.0cm}
 \caption{GRB as cosmological probes of the WHIM structures.
   The simulation  refers to an observation by ESTREMO with a
   microcalorimeter with 2 eV resolution below 1 keV
of a bright afterglow with a fluence of $10^{-5}$erg cm$^{-2}$
from 60 s to 60 ksec after the burst.}
 \label{fig_whim}
\end{figure}

\section{X-ray spectroscopy and GRB as cosmological probes}

 GRB
are the brightest and most distant sources in the Universe. The
radiation intensity of GRB's  is so high that they can be
detectable out to much larger distances than those of the most
luminous quasars or galaxies observed so far. In addition, long
bursts are now unquestionably associated with explosions of massive
stars, taking place in star formation regions. Taking advantage of
these GRBs properties, with ESTREMO we can address three key
themes of modern Cosmology and astrophysics:
\begin{itemize}
\item {\bf Discover and study the first "X-ray light" from
primordial gravitationally bounded object in the Universe at
z=10-30}. Our observational window on the Universe extends in
distance up to z=6-7, the reshift of the most distant object
discovered so far, and then recovers at z=1000, the epoch of
primordial fluctuations measured  by BOOMERANG and WMAP. The
formation of the first objects, stars, and protogalaxies, should
have taken place at epochs corresponding to z=10-30, certainly
beyond z=5. These first gravitationally bound proto-systems are
the result of the evolution of the primordial fluctuations
observed at z=1000, this evolution depending on cosmological
models and dark-matter properties. The large observational gap in
between these epochs is then particularly serious.  These objects
are obscure in the optical due to the dusty environment (and Ly
$\alpha$ forest absorption at z $>$ 5). On the other hand, X-rays
and gamma-ray photons produced by a GRB will easily pierce through
this environment, pin-pointing the location of the host galaxy and
allowing to measure its distance in X-rays by measuring the
redshift of X-ray features (X-ray redshift).

\item {\bf Trace the cosmic dark matter web at z $<$ 2 in X-rays}. In
the local (z $<$ 1) Universe the evolution of large-scale
structures dominated by dark matter is challenging the observers.
The sudden decrease of the baryon density in the local Universe is
one of the unresolved issues of Cosmology. The most intriguing
solution is that most of the baryon are in a hot phase, that can
be detected primarily through X-ray measurements. Numerical
simulations predict that,  at z$<1$, most of the baryons fall
  onto the cosmic web pattern of the dark matter, and are heated at
  $T\approx 10^6$K
  by shock mechanisms,
  forming filamentary and sheet-like structures \cite{ref:cen}. Such gas
  is called Warm-Hot Intergalactic Medium
  (WHIM). One of the most promising methods
 to detect and study  this component is by searching for the narrow absorption
  features -- the strongest of which is  OVII (at 0.574 keV in the rest
  frame) -- imprinted by the WHIM on the X-ray spectrum of a bright background
  object. Using bright GRB afterglows as background sources gives the big
advantage, with respect to AGN \cite{ref:nicastro}, to reach out
much larger distances, increasing the number of filaments through
the line of sight. For a burst at $z\gta 0.5$ at least one system
with an equivalent width $\gta 0.4$ eV is expected along a random
line of sight, while many more (8) are expected for just twice
weaker systems \cite{ref:hellsten}. In figure 1 we show one
expected absorption spectrum where, for clarity, we have limited
the number of weaker absorption lines to 3 plus one stronger line.

\item {\bf Study the history of metal enrichment in the Universe
from early epoch to the local Universe}.  X-ray "light" emitted
from distant GRB will be selectively absorbed at specific
frequencies by metals at the source, thus allowing to build up a
"map" of metal abundances and hence star formation rate as
function of the redshift.

\end{itemize}

\section{X-ray polarimetry and GRB engine}
Magnetic Fields play a major role in current theories of Gamma-Ray
Bursts. While energetics can be directly probed through multi-band
evolution of spectra, our knowledge of the structure of the
magnetic field and its genesis are totally committed to very
indirect evidences. Polarization would give a straightforward
evidence of the structure of the magnetic field and of its
evolution, enlightening the shock mechanism. Polarizations of the
order of 1\% have been detected in the optical band, in the
afterglow phase on timescales of 1 day. A polarization of $ \gta
60 $ \% was claimed\cite{ref:cob} in the prompt emission of a
burst (GRB021206). While this result is still to be confirmed, it
generated theoretical predictions on the role of a highly oriented
magnetic field, to suport the theory of magnetic fireballs
\cite{ref:lyu}. The study of how the polarization evolves (in
degree and angle) from the main burst to the afterglow phase in
the first minutes/hours can give a deep insight on the nature and
structure of the magnetic field and, possibly give important
constraints on the progenitor models. Gruzinov and Waxman
\cite{ref:WG}, by assuming the fireball to be composed of large,
mutually not ordered fragments of internally well ordered field
had predicted a high polarization level in the early phase,
quickly decreasing with the decrease of the Lorentz Factor. Other
authors \cite{ref:rossi,ref:sari} have explored the hypothesis
that in a GRB jet the magnetic fields are compensating to produce
a reduced polarization. When the relativistic  beaming cone
crosses the jet cone (namely close to the breaking of the light
curve) the symmetry is broken and a strong polarization would
appear for a limited time.  In Tab.1 we show the performances of
the ESTREMO mission, based on the X-ray Micro Pattern Gas Chamber
polarimeter \cite{ref:costa}. We choose a prudent configuration
for the detector so that meaningful measurements can be performed
only on bright bursts. From development studies we expect a
significant improvement of this figure. The polarimeter is
operated in alternative to the microcalorimeter. Nevertheless it
provides a good imaging and timing of the source, and energy
resolved polarimetry.

%\begin{figure}
%\includegraphics{f1e-5_r4_whim.ps}     % includes figure foo.eps
%\caption{Foo onteracting with bar}
%\end{figure}

\begin{table}
 \caption{Minimum polarization detectable with ESTREMO at
3$\sigma$ confidence for two bright GRBs and for two
representative bursting episodes of a Soft Gamma Repeater}
 \label{tab:polar}
  \begin{narrowtabular}{2cm}{cccc}
Source & $T_{Start}-T_{End}$ & MDP(\%) & Exposure (s)\\
 \hline
GRB970228 & 35-100 s   & 13 & 65 \\
GRB970228 & 100-300 s   & 13 & 200 \\
GRB990123 & 35-100 s  &    4  & 65  \\
GRB990123 & 100-300 s &   4  &  200  \\
\hline
SGR1900+14(normal burst)  & 35-1000 s &   20 & 965 \\
SGR1900+14(giant burst) & 35-1000 s  &  1  & 965  \\
    \hline
  \end{narrowtabular}
\end{table}

\acknowledgments


\begin{thebibliography}{0}
\bibitem{ref:cen} \BY{Cen~R. \atque Ostriker~J.P.}
\IN{Ap.J.}{514}{1999}{1}
\bibitem{ref:nicastro} \BY {Nicastro~F. et al}
\IN{Nature}{433}{2005}{495}
\bibitem{ref:hellsten} \BY{Hellsten~U., Gnedin~N.Y \atque Miralda-Escude~J.}
\IN{Ap.J.}{505}{1998}{56}
\bibitem{ref:cob} \BY{Coburn~W. \atque Boggs~S.L.}
  \IN{Nature}{423}{2003}{415}
\bibitem{ref:lyu} \BY{Lyutikov~M., Pariev~V.I. \atque Blandford~R.D.}
  \IN{Ap.J.}{597}{2003}{998}
  \bibitem{ref:WG} \BY{Gruzinov~A. \atque Waxman~E.}
  \IN{Ap.J.}{511}{1999}{852}
\bibitem{ref:rossi} \BY{Rossi~E., Lazzati~D., Salmonson~J.D. \atque Ghisellini~G.}
  \IN{M.N.R.A.S.}{354}{2004}{86}
\bibitem{ref:sari} \BY{Sari~R.}
  \IN{Ap.J.Let.}{524}{1999}{L43}
\bibitem{ref:costa} \BY{Costa~E. et al.}
  \IN{Nature}{411}{2002}{662}
  %\bibitem{ref:apo} \BY{Boccaccio~G. \atque de~Cam\~oes~L.}
%  \IN{Phys. Rev. A}{13}{1999}{12};
%  \SAME{69}{999}{1666}.
%\bibitem{ref:pul} \BY{Pulci~L.}
%  preprint INFN 8181.
%\bibitem{ref:bra} \BY{Bragg~B.}
%  \TITLE{Tender comrade},
%  in \TITLE{Workers Playtime},
%                  edited by \NAME{Tizio A. \atque Caio B.}
%                  (Unexeditor, Bologna) 1997, pp.~1-10.

\end{thebibliography}
\end{document}